\documentclass[a4paper,fleqn]{cas-dc}

\usepackage{hyperref}
\usepackage{siunitx}
\usepackage{lineno} % for line numbering
\usepackage{graphicx}%
\usepackage{multirow}%
\usepackage{amsmath,amssymb,amsfonts}%
\usepackage{amsthm}%
\usepackage{mathrsfs}%
\usepackage[title]{appendix}%
\usepackage{xcolor}%
\usepackage{textcomp}%
\usepackage{manyfoot}%
\usepackage{booktabs}%
\usepackage{algorithm}%
\usepackage{algorithmicx}%
\usepackage{algpseudocode}%
\usepackage{listings}%
\usepackage{physics} % to write derivatives
\usepackage[separate-uncertainty = true,multi-part-units=single]{siunitx} % uncertainties

\usepackage[authoryear,longnamesfirst]{natbib}
\def\tsc#1{\csdef{#1}{\textsc{\lowercase{#1}}\xspace}}
\tsc{WGM}
\tsc{QE}
\begin{document}

\let\WriteBookmarks\relax
\def\floatpagepagefraction{1}
\def\textpagefraction{.001}

\shorttitle{Gravity-Induced Ice Compaction and Subsurface Porosity on Icy Moons}
\shortauthors{C. Mergny and F. Schmidt}
\title[mode=title]{Gravity-Induced Ice Compaction and Subsurface Porosity on Icy Moons}

% First Author
\author[1]{Cyril Mergny}[orcid=0009-0002-1910-6991]
\cormark[1]
\ead{cyril.mergny@universite-paris-saclay.fr}

% Second Author, etc...
\author[1, 2]{Frédéric Schmidt}

% Affiliations
\affiliation[1]{
    organization={Université Paris-Saclay, CNRS, GEOPS},
    city={Orsay},
    postcode={91405}, 
    country={France}}
    
\affiliation[2]{
    organization={Institut Universitaire de France},
    city={Paris},
    country={France}}

%\linenumbers
%\doublespacing

\begin{abstract}
Our understanding of the surface porosity of icy moons and its evolution with depth remains limited, including the precise scale at which ice compaction occurs under self-weight pressure.
This parameter is of crucial interest for the correct interpretation of current remote sensing data (spectroscopy in the visible, infrared to passive microwave) but also for planetary exploration when designing a lander, a rover or a cryobot. In situ exploration of the ice crust would require knowledge about subsurface porosity.
This study employs a compaction model solely driven by overburden pressure based on prior research. The formulation for density as a function of depth, incorporates an essential parameter: the ice compaction coefficient. 
To determine this coefficient, we fit our depth-dependent density model to existing data obtained from Earth-based measurements of ice cores in Antarctica and North Greenland.
Our results yield a typical lengthscale for ice compaction on Earth of approximately $ \SI{20.1(6)}{m}$ , consistent with the existing literature.
We apply the model to Europa, which due to its lower gravity, has a typical ice compaction scale of $\SI{150(4)}{m}$, when assuming an Earth-like compaction coefficient. We compare it with the depths scanned by current spaceborne data and find that porosity can be considered constant when accounting only for gravity-induced compaction.
\end{abstract}

% Research highlights
%\begin{highlights}
%\item 
%\item 
%\item 
%\end{highlights}


\maketitle

\section{Introduction}

Icy moons, characterized by an icy crust at the surface, present unresolved questions regarding the microscopic properties of their surface and subsurface. 
Various studies provide compelling evidence that the surface of icy satellites is made of a porous material.
One significant piece of evidence comes from brightness temperature measurements, where thermal models obtain a thermal inertia significantly smaller than that of solid ice \citep{Spencer1989, Rathbun2010}. The most plausible explanation for this reduction in thermal inertia is to introduce porosity into the thermal properties, as it changes the density and conductivity of ice \citep{Ferrari2016}. 
Other indications are found in spectroscopy measurements of icy moons' surfaces showing improved fits when incorporating a porosity factor \citep{Mishra2021, Mermy2023} or from the potential existence of O2 molecules trapped in bubbles within Europa's near-surface ice \citep{Oza2018}, indicating the presence of a porous regolith.
Various processes may contribute to the development of a porous surface, including the deposition of snow onto the surface or space weathering, which could potentially transform initially solidified ice into a porous material through constant bombardment by micrometeorites and particles.

Specifically, the exact texture of the near-surface remains unclear. Is the ice made of grains with high porosity that compact with increasing depth? Of a granular layer overlaying a denser slab? Or of any other possible configuration?
Subsurface porosity is unknown but it constitutes a crucial parameter when building subsurface models for the analysis of spaceborne data, whether aiming to derive thermal properties or optical properties. 
A common assumption in remote sensing analysis, is that the probed material has a unique thermal inertia \citep{Spencer1989}. This implies that the subsurface is homogeneous, thus having constant porosity through all the depths probed. 
On Europa, the diurnal skin depth is estimated to be on the order of few centimeters \citep{Ferrari2016}, while the seasonal skin depth may extend to dozens of meters. Hence it is unclear whether Europa porosity changes at the depths probed by brightness temperature retrievals.

On Earth at these depths, notable changes in porosity from ice core measurements have been observed on Antarctica and North Greenland ice, with variations spanning approximately 20 meters \citep{Alley1982, Horhold2011, Gerland1999}.
Earlier work \citep{Herron1980, Alley1982}, identified critical density thresholds, such as $\SI{550}{kg.m^{-3}}$ for the transition from snow to firn and $\SI{840}{kg.m^{-3}}$ from firn to coarse firn, resulting in the proposal of three distinct empirical models to describe the density transitions within the snow structure.
However, a more recent investigation \citep{Horhold2011}, using a comprehensive measurements of 16 ice cores Antarctica and North Greenland did not reveal clear transitions in the ice density profile.
This finding suggests the possibility of a unified formulation to describe the entire density variations in these ice cores.
Other efforts have been made to model snow compaction more realistically on Earth \citep{Cuffey2010}, focusing on complex interactions involving pressure, temperature and sintering. These models  \citep{Wilkinson1988, KopystynSki1993, Arnaud2000, Meyer2020} often rely on the fundamental sintering equations first formulated by Swinkels and Ashby \citep{Swinkels1981}.

On Europa, studies \citep{Nimmo2003, Johnson2017, Howell2021} have primarily addressed the dynamic variations of porosity over time within the whole ice shell, predominantly attributed to ice viscosity.
However, these studies tend to concentrate on deeper depths ($\sim$ 10s of km), where accurate modeling of the shallow subsurface ($<\sim$10s m) is not required.
While exploring other compaction-related processes like sputtering, as observed in some planetary science studies \citep{Raut2008, Schaible2017}, and late-stage sintering \citep{Molaro2019}, holds promise for future research, our current focus remains on modeling ice compaction under the influence of overburden pressure due to gravity.
Hence we address a fundamental question: What is the characteristic lengthscale of ice compaction on icy moons when considering solely the overburden pressure induced by gravity?

The main objective of this study is not to determine the precise surface porosity of icy moons, as the existing data currently available makes it challenging to constrain this parameter.
Rather, our focus is to determine what would be the compaction profile of ice for any value of surface porosity. 
We first adapt the approach undertaken in \citep{Wilson1994} for the compaction of silicate rock on Earth, Mars and Venus to model the compaction processes of ice on icy worlds. Then, we propose an analogy with a viscous model  initially proposed for magma \citep{Fowler1985} and used by the planetary science community \citep{Nimmo2003, Johnson2017} to study the ice shell. The model will be tested on Earth cases and discussed for icy moons, with a particular focus on Europa.

\section{Methods}

As pointed out by Earth based studies \citep{Mellor1977}, densification of snow under static loading remains unclear.
On the one hand, compressive load result in depth-density relations that are almost time-invariant as stated by Sorge's Law \citep{Bader1954, Bader1960}, implying that an elastic model would be relevant.
On the other hand, snow is known to creep, thus densification can be treated as a continuous time-dependent process, using a viscous flow formulation.

\subsection{Compaction Model}
\subsubsection{Generalized Porosity Expression}

Following \citep{Wilson1994}, a continuous density function can be used to model the variations of ice density $\rho$ with depth $z$  when compaction of the pore space is solely induced by the overburden pressure.
The assumption made by the authors \citep{Wilson1994} is that the porosity $\phi$ shows an exponential decay with increasing pressure $P$, leading to the form:
\begin{equation}
\phi(z, t) = \phi_0  \exp\left(-\lambda(z,t) P(z)\right)
    \label{eq:phiz}
\end{equation}
where $\phi_0$ is the top surface porosity value and $\lambda(z, t)$ will be called the compaction coefficient of the considered material.
The compaction coefficient, due to its unit in $\mathrm{Pa}^{-1}$, can also be thought as the inverse of the characteristic pressure required for substantial material compaction.
Unlike Wilson's approach, this generalized form uses a  compaction coefficient $\lambda(z, t)$ which may not be constant with depth or time. This anticipates that factors such as temperature, porosity, compaction time and others can influence the compaction coefficient, thereby affecting the porosity profile $\phi(z, t)$.\\

\subsubsection{Density Profile for a Constant Compaction Coefficient}
To derive the density profile expression, we follow Wilson's assumption \citep{Wilson1994} of a constant compaction coefficient $\lambda$, with respect to depth.
Although the limitations of such an assumption are discussed in \ref{sec:depth_visco}, it is necessary in order to obtain a simplified expression of the compaction lengthscale.

It follows from the definitions of the bulk density $\rho_{\mathrm{b}}$ and pore space that the density at any given depth is given by the relation:
\begin{equation}
    \rho(z) = \rho_{\mathrm{b}} (1-\phi(z))
    \label{eq:rhoz}
\end{equation}
Since the pressure at depth $z$ is given by the weight of the above ice layers, an increase of $\dd{z}$ will increase $\dd{P}$ by the relation
\begin{equation}
    \dd{P(z)} = \rho(z) g \dd{z}
    \label{eq:dPdz}
\end{equation}
where $g$ is the gravity acceleration at the considered body's surface.

Combining Equations \eqref{eq:phiz}, \eqref{eq:rhoz} and \eqref{eq:dPdz}, leads to a differential equation for $P(z)$, which after separation of variables and integration takes the form
\begin{equation}
    P(z) = \frac{1}{\lambda} \ln \left( \phi_0 + (1-\phi_0) \exp\left(\dfrac{z}{H} \right)  \right).
    \label{eq:P_wilson}
\end{equation}
and provides the expression for density:
\begin{equation}
    \rho(z) = {\rho_{\mathrm{b}}} \left(1 + \frac{\phi_0}{1-\phi_0} \exp\left(-\dfrac{z}{H} \right)  \right)^{-1}
    \label{eq:density_wilson}
\end{equation}
with $H$ the characteristic lengthscale of compaction given by
\begin{equation}
    H = \dfrac{1}{ \lambda g \rho_{\mathrm{b}}  }.
    \label{eq:lengthscale}
\end{equation}
Using this formulation, the depth-density profile of an icy layer can be specified in terms of the three model parameters: the top surface porosity $\phi_0$, the ice compaction coefficient $\lambda$ and the gravitational acceleration of the target body $g$.

\subsection{Elastic and Viscous Compaction}
The assumption of an exponential decay of porosity $\phi(z)$ with rising pressure $P(z)$ made by \citep{Wilson1994}, was revisited in Appendix \ref{sec:appendix_rheo}. We demonstrated that it can be mathematically derived from both the stress-strain rate relationship of a purely elastic material or of a purely viscous material.  
The combination of high tidal heating and low gravity of many outer planet satellites leads to conditions where pore close more readily viscously due to high temperatures rather than due only to self gravity.
Hence in many studies of Europa's ice shell \citep{Nimmo2003, Howell2021}, ice compaction is approached by considering it as the compression of a viscous flow, using Fowler's law of porosity \citep{Fowler1985}:
\begin{equation}
    \dfrac{\partial{\phi}}{\partial{t}}(z,t) = - \phi(z,t) \frac{P(z)}{\eta(z,t)}
\end{equation}
where $t$ is the time of compaction, and $\eta(z, t)$ the viscosity of the considered material at a specific depth and time.
This differential equation can be derived from the stress-strain rate relationship of a purely viscous material, as demonstrated in Appendix \ref{sec:appendix_visc}.
This leads to the expression of porosity:
\begin{equation}
    \phi(z, t) = \phi_i(z) \exp(-\frac{t}{\eta(z,t)} P(z))
    \label{eq:phi_fowler}
\end{equation}
where $\phi_i(z)$ is the initial porosity profile at $t=0$.

It is unclear whether near-surface ice on icy bodies undergoes elastic compaction, viscous creep, or a simultaneous combination of both processes. 
Determining the dominant compaction mechanism is challenging, primarily due to the poorly constrained viscosity of near-surface porous ice at these conditions, which can vary on multiple orders of magnitude. 
The main focus of this study is not to quantify which compaction process dominates on icy moons, but  rather to understand how each process would behave independently and can be related to single parameter: the compaction coefficient.
Elastic compaction and viscous flow are two distinct compaction processes but derivation of compaction from both elastic or viscous processes result in the same general mathematical form of the porosity given by Equation \eqref{eq:phiz}.
The elastic derivation shows a direct inverse relationship between the compaction coefficient and the ice Young's modulus $E$:
\begin{equation}
    \lambda \Leftrightarrow \frac{1}{E}.
\end{equation}
while the viscous flow derivation leads to a relationship between the compaction coefficient $\lambda$, the compaction time $t$, and the viscosity $\eta$:\begin{equation}
    \lambda(z, t) \Leftrightarrow \dfrac{t}{\eta(z, t)}.
    \label{eq:lambda_t_eta}
\end{equation}
The relationship of Equation \eqref{eq:lambda_t_eta} elucidates that selecting a value for $\lambda$ on any icy body imposes constraints on a parameter pair: viscosity and compaction time.
If we adhere to Wilson's analogy method \citep{Wilson1994}, we must assume that the compaction coefficient for a planetary ice body is comparable to the value observed on Earth. 
However, if the compaction coefficient deviates from the Earth's measurement, Equation \eqref{eq:lambda_t_eta} provides the relationship between $\lambda$, $t$, and $\eta$.

\section{Results}

Number of reasons may lead to changes in the values of the compaction coefficient of ice between Earth and icy moons.
The significantly lower temperature, lack of atmosphere and the different environment found on the icy moons  could change how ice reacts to compression. 
To the best of our knowledge, we currently lack a method to properly obtain the compaction coefficient on icy moons. 
Therefore,  in this section we will first describe how ice would compact due to the lower gravity on icy moons assuming the compaction coefficient is the same as on Earth. 
Then we explore how other compaction mechanisms like viscous creep may come into play, effectively changing the compaction coefficient.

\subsection{Constant Earth-like Compaction Coefficient}
\label{sec:ela_result}

\begin{figure*}
	\centering
	  \includegraphics{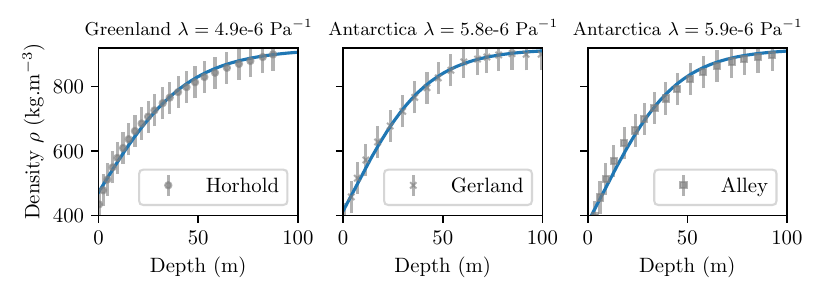} 
    \caption{Ice depth-density profiles for three ice cores  (\textit{Left}) North Greenland ($77.3$ S,$ -49.2$ W), (\textit{Middle}) Berkner Island, Antarctica A1 ($79.3$ S, $45.4$ W), (\textit{Right}) Ridge BC, Antarctica A2 ($82.5$ S, $136.4$ W). Grey dots depict measured density data with associated error bars \citep{Horhold2011, Gerland1999, Alley1982}, while blue lines represent the best-fit density function using Equation \eqref{eq:density_wilson}. }
    \label{fig:fit_compact}
\end{figure*}

To determine the value of the ice compaction coefficient on Earth, an analysis of density measurements with depth derived from Earth-based data is conducted.
This analysis uses a dataset comprising measurements from ice cores located in North Greenland ($77.3$ S, $-49.2$ W) \citep{Horhold2011}, Berkner Island, Antarctica A1 ($79.3$ S, $45.4$ W) \citep{Gerland1999}, and Ridge BC, Antarctica A2 ($82.5$ S, $136.4$ W) \citep{Alley1982}.
%To determine the compaction coefficient for Earth ice, we have analysed density measurements with depth obtained from Earth-based data. 
While sintering and melting occur on Earth, on large ice sheet such as Greenland and Antarctica, they are only confined in the first few meters depths \citep{ Alley1982}. Therefore, we conclude that these processes do not significantly impact the density variations spanning dozens of meters observed in these ice cores.
Knowing the bulk ice density $\rho_{\mathrm{b}} = \SI{917}{kg.m^{-3}}$, we have fitted the depth density profile from Equation \eqref{eq:rhoz} to the three ice cores shown in Figure \ref{fig:fit_compact}.
For example, for the B26 North Greenland ice core, the best fit results in an initial surface porosity $\phi_0 = 48.5 \pm 0.7 \%$ and an ice compaction coefficient of $\lambda = \SI{4.9(1.5)e-6}{Pa^{-1}}$.
The average value  over the three locations studied here is $\overline{\lambda} =  \SI{5.5(2)e-6}{Pa^{-1}}$ and will be used as a reference for the compaction of ice on Earth for the remaining of this article.

\begin{figure}[htbp]
    \centering
    \includegraphics[width=0.9\linewidth]{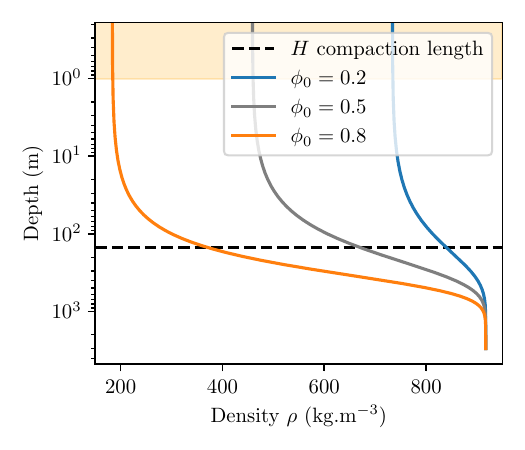} 
    \caption{Modeled ice densities as  function of depth for Europa's ice shell and different starting porosities using Earth's compaction coefficient $\overline{\lambda} = \SI{5.5(2)e-6}{Pa^{-1}}$. For reference, the characteristic compaction length $H$ for Europa is represented by the dotted black line. The orange zone shows the surface layers affected by solar temperature variations, called the seasonal thermal skin depth. We observe that porosity remains nearly constant around the thermal skin depth, due to Europa's low gravity. }
    \label{fig:density_depth}
\end{figure}

Having obtained Earth's ice compaction coefficient $\overline{\lambda}$, we extend our analysis by applying the density-depth formulation presented in Equation \eqref{eq:rhoz} to other planetary surfaces.
While this method could be applied to any icy moon, our focus lies on Europa, where the surface gravitational acceleration is estimated to be $g = \SI{1.315}{m.s^{-2}}$ \citep{Anderson1998}. 
For any given surface porosity, the formulation of Equation \eqref{eq:density_wilson} enables us to determine the density profile at various depths, accounting only for self-weight-induced ice compaction.
In Figure \ref{fig:density_depth}, we present distinct density profiles for a range of initial porosities on Europa for an ice compaction coefficient $\overline{\lambda}$. 
It is evident that, regardless of the starting porosity, significant compaction effects are not noticeable until reaching depths deeper than 100 meters. This outcome aligns with the prediction from Equation \eqref{eq:lengthscale}, which illustrates that the compaction lengthscale is largely independent of the initial porosity, as shown in Figure \ref{fig:density_depth}.

When evaluating this expression for Europa, we derive a compaction lengthscale of $H = 150 \pm \SI{4}{m}$.  At the subsurface, for example for depths under 5 m, the porosity change ranges from approximately $2\%$ to less than $1\%$, depending on the initial porosity $\phi_0$. 
This result aligns with radar reflectivity models, which suggest that Europa has a surface layer with appreciable porosity that is at least meters thick  \citep{Johnson2017}.
Visible compaction effects only become prominent at greater depths, near the compaction lengthscale $H$.

\begin{table*}[width=.9\textwidth,cols=4,pos=h]
    \centering
    \begin{tabular}{lccc}
        \toprule
        Name & Radius ($\SI{}{km}$) & Surface Gravity ($\SI{}{m.s^{-2}}$) & Gravity-Induced Compaction Length $H$ ($\SI{}{m}$) \\
        \midrule
         Earth & 6370 & 9.81  & 20 \\
         Ganymede & 2630 & 1.43  & 140 \\
         Europa & 1560   &  1.32 & 150 \\
         Callisto & 2410  & 1.24 & 160 \\
         Pluto &  1190 &  0.62 & 320 \\ 
         Iapetus &  1470 &  0.22 & 880 \\
         Enceladus & 250  & 0.11 & 1750 \\
        \bottomrule
    \end{tabular}
    \caption{Ice compaction lengthscales under gravity-induced pressure for different bodies in the Solar System assuming a constant Earth-like compaction coefficient of ice $\overline{\lambda} = \SI{5.5(2)e-6}{Pa^{-1}}$. Note that other processes like sputtering or sintering, may also compact the ice leaving to different compaction lengths.}
    \label{tab:icy_moons}
\end{table*}

We have applied the same analysis for various icy bodies in the Solar System and present in Table \ref{tab:icy_moons} their compaction lengthscale.
We found that the lengthscales of gravity-induced compaction range from $\sim \SI{20}{m}$ to $\sim \SI{2000}{m}$ depending on gravity and thus size of the bodies.
In general, due to a lower gravitational acceleration, ice compaction lengthscales are considerably higher than those observed on Earth.
While our study focused on Solar System icy bodies, it anticipates that exoplanetary icy bodies, potentially discovered in the future, would benefit from these findings in interpreting surface properties through direct imaging \citep{Hu2012, Berdyugina2019}.

\subsection{Viscous Creep Scenario}

The previous analysis proposed to use the compaction of Earth ice as an analog to the compaction of ice on an icy moons.
However it is possible that the compaction coefficient $\lambda$ for a given planetary ice has a different value than the one measured on Earth $\overline{\lambda}$.
If the ice composition or its physical properties like viscosity are different than the one on Earth, the viscous compression model would be helpful to extend the previous work. 
Also, while the measurements from Earth ice cores have allowed particular time for ice to compact, we remain uncertain about the  compaction time on icy moons.

\begin{figure}
    \centering
    \includegraphics[width=1.05\linewidth]{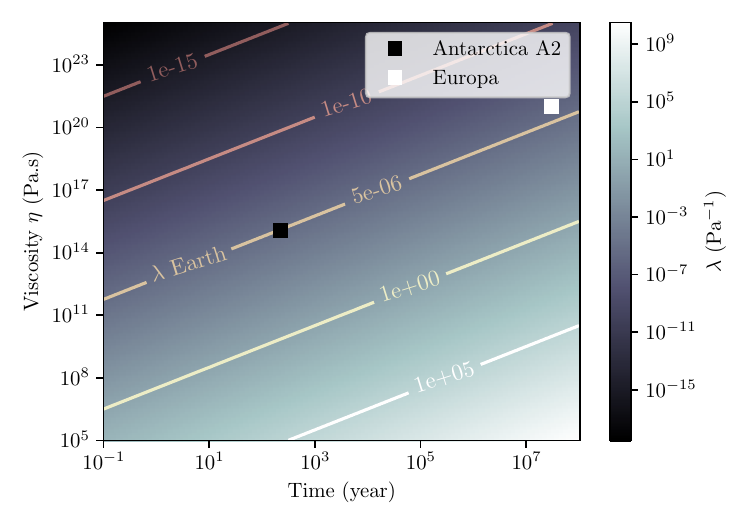} 
    \caption{Evolution of the compaction coefficient $\lambda(\eta, t)$ with the ice viscosity and the compaction time. The average ice compaction coefficient measured on Earth is given by the yellow line $\overline{\lambda} = \SI{5.5e-6}{Pa^{-1}}$. Black square represents the location of the Antarctica A2 dataset in this graph, estimated from the local accumulation rate of snow. White square represents our scenario for Europa, where the compaction time is taken to be the ice crust age and for a viscosity of  $\eta = \SI{e21}{Pa.s}$ \citep{JaraOrue2011}. }
    \label{fig:lambda_vs}
\end{figure}

In this model, we can verify that the compaction coefficient $\lambda$ is coherent with well known values of ice viscosity on Earth.
Since glaciers are formed by accumulation at the top, we propose to relate the elapsed time of compaction at depth $z$ to the accumulation rate of snow $A$. To achieve this, we use a simplified expression from \citep{Bader1954, Bader1960}:
\begin{equation}
    t(z) = \dfrac{z}{A}.
    \label{eq:accumulation}
\end{equation}
For the Antarctica A2 ice core, a snow accumulation rate of $A=\SI{100}{mm.yr^{-1}}$ was estimated \citep{Alley1982}. Applying Equation \eqref{eq:accumulation} to the characteristic compaction depth $z = H=\SI{19}{m}$ results in a compaction time of $t=\SI{190}{yr}$. 
Then, from the compaction coefficient of the A2 ice core, we can use Equation \eqref{eq:lambda_t_eta} to estimate the viscosity of the compacted snow in this region as $\eta=\SI{1.0e15}{Pa.s}$ (see Figure \ref{fig:lambda_vs}, black square). This viscosity value is compatible with the compactive viscosity mentioned by \citep{Mellor1974} for Antarctica ice, falling within the range of $\SI{e14}-\SI{e18}{Pa.s}$. The viscous ice model is compatible with this data.\\

In order to apply this model to Europa, we first need to establish a compaction scenario. 
While an accurate accumulation rate remains to be determined for Europa, we propose considering the compaction time to be the ice crust age, estimated at approximately $t=\SI{30}{Myr}$  \citep{Zahnle2003}. Using the surface age in this way provides an upper boundary value for compaction time.

The viscosity of Europa's subsurface ice remains unknown. It is challenging to model notably due to its high dependency on various parameters including temperature, stress, porosity  and grain size.
Here we first present a simple analytical form to estimate the viscosity of ice, then in Section \ref{sec:depth_visco} we discuss the relevance and limitations of such approach.
Studies focusing on the deeper portion of the ice shell use a temperature-dependent viscosity law \citep{Thomas1986, Nimmo2003}:
\begin{equation}
    \eta(z) = \eta_b \exp \left( \frac{Q}{R} \left[ \frac{1}{T(z)} - \frac{1}{T_b} \right] \right)
    \label{eq:visco_temp_law}
\end{equation}
where $\eta_b$ and $T_b$ are the viscosity and temperature at the base of the ice shell, $Q$ is an activation energy and $R$ the gas constant.
While this viscosity law is suitable for modeling the base of the ice shell, near the surface, Europa's low temperature $\sim \SI{100}{K}$ results in unrealistically high ice viscosity values exceeding $\SI{e26}{Pa.s}$. As highlighted by \citep{Thomas1986}, appropriately modelling the subsurface would require a warmer surface to counterbalance these excessively high viscosity.

To the best of our knowledge, there has not been a comprehensive study modeling the shallow subsurface viscosity of Europa. 
However, various authors \citep{JaraOrue2011, Kihoulou2021} have proposed the use of a cut-off value for viscosity to address the exceptionally high values given by Equation \eqref{eq:visco_temp_law}. Following the suggestion by \citep{JaraOrue2011} for Europa's lithosphere, we consider $\eta = \SI{e21}{Pa.s}$ as a potential viscosity for Europa's icy subsurface.
Using Equation \eqref{eq:lambda_t_eta}, we find that ice with such viscosity, compacted over $30$ million years, yields a compaction coefficient value of $\lambda = \SI{0.9e-6}{Pa^{-1}}$ and a compaction lengthscale of $H = \SI{880} {m}$.
It is noteworthy that the compaction coefficient value, $\lambda$, although falling within the large potential range of values given by Equation \ref{eq:lambda_t_eta}, is of the same order of magnitude as Earth's value, as illustrated in Figure (\ref{fig:lambda_vs}, white square).

Local-scale processes, such as the formation of chaos terrain or cryovolcanism on Europa, could result in more recent and locally warmer ice, thereby changing the compaction lengthscale.

\section{Discussion}

\subsection{Depth and Time Dependant Viscosity}
\label{sec:depth_visco}
The viscosity of ice depends on various factors such as temperature, porosity, stress, and grain size.
Across the depths considered in this study, these parameters may vary with depth (e.g., temperature due to a thermal gradient) or with time (e.g., porosity due to viscous pore closure). 
Our model, primarily focused on the near surface of icy moons, assumes a constant viscosity with depth and time. Here, we discuss both the relevance and limitations of this approach.

\subsubsection{Porosity Dependant Viscosity}
We expect the surface of icy Galilean moons to exhibit high porosity \citep{Black2001, Johnson2017} and studies on porous media \citep{Mackenzie1950, Sura1990} suggest that the viscosity decreases non-linearly with increasing porosity.
Theses findings imply that the viscosity of the porous subsurface ice on Europa would be significantly lower than the viscosity of bulk ice.
In fact, porous snow on Earth, with a viscosity in the range of $\SI{e4}-\SI{e9}{Pa.s}$ \citep{Mellor1974, Bartelt2000, Camponovo2001}, can be orders of magnitude less viscous than compacted ice at Earth's temperatures, which has a viscosity around $\SI{e12}{Pa.s}$ \citep{Fowler1997}.

Assuming that the viscosity changes with depth would be a more realistic approach but it would require solving both the porosity and viscosity evolution as an iterative and coupled process.
While this would be valuable for future research, the primary goal in this study is to present a simple yet analytical form that helps making comparative planetology.

Elastic compaction tends to have porosity that decreases exponentially with depth \citep{Clifford1993, Clifford2001, Wieczorek2013}.
This can be intuited from Equation \eqref{eq:phiz}, when considering a constant compaction coefficient with depth.
Although Equation \eqref{eq:phi_fowler} has an exponential form, the porosity profile with depth does not necessary follow an exponential curve. 
Indeed,  for deeper portion of the ice shell the viscosity is the primary driver of ice compaction, and not the overburden pressure.
Hence viscous flow tends to have a relatively constant porosity until the temperature is great enough at depth to rapidly close porosity in a geologically short period of time \citep{Besserer2013, Wieczorek2013, Gyalay2020}.

\subsubsection{Variations of Temperature with Depth}

For modeling the compaction of near-surface ice, our model assumes a constant temperature, though temperatures may fluctuate within these depths due to a thermal gradient. 
Here, we quantify the variations of temperatures in the near-surface temperatures and explore their impact on viscosity.
The temperature profile in the model by \citep{Nimmo2003} assumes a mean surface temperature of 100 K for Europa, with results indicating a constant temperature within the first kilometer.
Similar results are observed at the near surface of Enceladus \citep{Besserer2013}. The authors \citep{Besserer2013} propose to use the improved model of conduction, where the changes of conductivity with porosity are taken into account \citep{Shoshany2002}. 
Even with this refined model, their results indicate that temperature variations with depth is less than 5 K within the first hundred meters. At these depths, tidal heating is negligible \citep{Tobie2003}. 
Although the ice viscosity is strongly dependent on temperature, minor temperature fluctuations in the near surface, here inferior to $5$ K, would only result in a change in viscosity of one order of magnitude.  
Hence in contrast with the substantial uncertainty in viscosity variations with porosity, temperature variations with depth near the surface do not induce significant viscosity changes.
We conclude that adding a comprehensive description of the thermal profile would remove the simplicity of the analytical form, without gaining substantial additional information.

\subsubsection{Constitutive Flow Laws}

An important aspect that has not been developed in our model is the rheology of ice under different stress and grain-size conditions. 
Goldsby and Kohlstedt \citep{Goldsby2001} have worked on formulating a constitutive equation for the effective viscosity \citep{Besserer2013} (or strain-rate) under different creep regimes:
\begin{equation}
    \eta(T, P, D) = \left[ \eta_{\mathrm{diff}}^{-1} + \eta_{\mathrm{disl}}^{-1} + (\eta_{\mathrm{bas}} + \eta_{\mathrm{gbs}})^{-1} \right]^{-1} 
\end{equation}
where $D$ is the grain size  and the subscripts diff, disl, bas, and gbs refers to different creep regimes, namely diffusion ﬂow, dislocation creep, basal slip, and grain boundary sliding.

\begin{figure}
    \centering
    \includegraphics[width=0.9\linewidth]{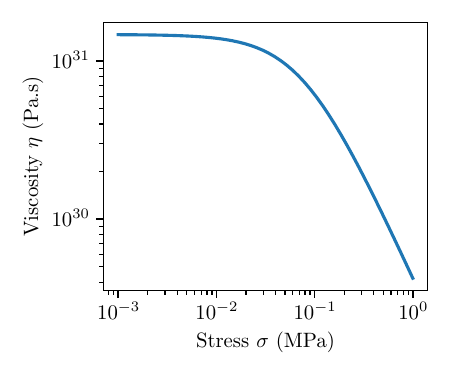} 
    \caption{Viscosity vs stress using Goldsby's \citep{Goldsby2001} constitutive equation and parameters  for a temperature of 100 K and a grain size of $\SI{100}{\mu m}$.}
    \label{fig:viscosity}
\end{figure}

Using the parameters from \citep{Goldsby2001} [see Tables 5 and 6] we found that at $100$ K,  the viscosity does not vary significantly within the range of stresses inferior to 0.1 MPa, typical of the near surface conditions on icy moons (see Figure \ref{fig:viscosity}).
However, interpreting these results is challenging, as the temperatures used to obtain the activation energies, around 250 K, are significantly higher than those on icy moons, around 100 K. 
Since the experimental data have been collected at temperatures around 250 K (viscosity around $\SI{e12}{Pa.s}$), it seems difficult to extrapolate these parameters  to icy moons cold environment (viscosity of $\SI{e31}{Pa.s}$ on Figure \ref{fig:viscosity} at 100 K). 
Conducting experiments to observe creep behavior in an environment with temperatures similar to those found on icy moons would be beneficial. However, it remains challenging due to the significant timescale for relaxation of ice at such low temperatures.

\subsection{Comparison with Cold Compaction Experiments}
A set of cold compaction experiments \citep{Durham2005} provided valuable insights into the compaction of cold ice at high pressure ($>$MPa) over relatively short timescales ($\sim$ 15h). 
As the authors acknowledged, densification can occur at all pressures, including those lower than the range covered in the experiments.
In our study, we have looked at the compaction of ice for lower pressure. At a depth of 100 meters, with, for example, a porous ice density of approximately $\SI{500}{ kg.m^{-3}}$, the overburden pressure is around $\SI{0.05}{MPa}$. 
Although we cannot precisely determine how ice would compact at these pressures, since it is outside the range of the experiments, their results at high pressure suggest that minimal elastic compaction would occur at lower pressure. 
Such experimental result would mean, that under elastic compression, the compaction coefficient of cold ice $T<100K$ is significantly lower than the one measured at Earth's temperature, $\overline{\lambda
}$. 
This would lead to a compaction lengthscale significantly higher than the one presented in Table \ref{tab:icy_moons}.
If anything, these experiments confirm that it is even more unlikely to see porosity variations at the depth probed by the current remote sensing instruments (except for the upcoming JUNO/MRW data, which may probe depths of dozens of kilometers.). 

Another result suggested  by these experiments, is that even after high pressure compaction,  substantial residual porosity  remains.
If this is indeed the case, then it would be necessary to change the expression of Equation \ref{eq:phiz} by adding a residual porosity term:
\begin{equation}
\phi(z) = \left( \phi_0 - \phi_{\infty} \right) \exp\left(-\lambda P(z)\right) + \phi_{\infty}
    \label{eq:phiz_durham}
\end{equation}
where $\phi_{\infty}$ is the residual porosity from elastic compaction.
Experiments on Earth snow \citep{Cuffey2010}, suggest that packing can not reduce porosity below 40\% and other mechanism must be responsible for further densification.
This porosity could be then be removed by viscous creep or by pressure sintering \citep{Cuffey2010} over long period of time (millions of years at 120 K).
The experiments from \citep{Durham2005} were conducted over relatively short timescales of about 15 hours, much shorter than the relaxation time of viscous ice at these temperatures.  In our study, we looked at the viscous creep of ice over timescales of 200 years on Earth and 30 millions years on Europa. We believe that changes in porosity due to viscous creep over such extended periods could not have been detected in Durham's experiments.

\section{Conclusion}

This study has used Earth ice core measurements to determine a fundamental parameter related to the compaction of ice under the influence of gravity: the ice compaction coefficient. As a result of this analysis, we were able to infer the typical gravity-induced compaction length on Europa of $H = \SI{150(4)}{m} $, assuming an Earth-like compaction coefficient of ice. By employing the same method, we determined that the compaction lengthscale for icy bodies within the Solar System spans a range from $\SI{20}{m}$ to $\SI{2000}{m}$.

The compaction lengthscale significantly exceeds the wavelengths and penetration depths within the visible and near-infrared ranges, which can extend up to only a few centimeters. 
Thermal infrared emissions allow to estimate features as deep as the diurnal and seasonal thermal skin depth, which is typically less than $\sim \SI{10}{cm}$ and less than dozen of meters for Europa, depending on the thermal diffusivity. At such depths, we expect that gravity-induced compaction will lead to relatively minor changes in porosity.
Still, recent observations from the JUNO MicroWave Radiometer could probe up to dozens of kilometers deep into the Galilean moons' surfaces \citep{Brown2023}. Incoming analysis of the observations for Europa would reveal new constraints on porosity of the ice shell.

The strength of the models presented in this article lies in their simplicity and analytical solutions. Enhancements can be made by considering the strong influence of porosity, temperature, stress and grain size on viscosity and developing a more precise compaction scenario for the crusts of icy moons. 
Additionally, other compaction mechanisms are likely to come into play near the surface, such as those induced by sputtering or sintering \citep{Schaible2017, Raut2008}, potentially observable by space instruments. The effect of these mechanisms on the ice porosity profile is not yet fully understood. While dry water ice metamorphism and sputtering-induced sintering have been studied for planetary surfaces \citep{Molaro2019, Schaible2017}, their quantitative connection to ice densification is yet to be established.

Given these results, upcoming missions should consider the use of large wavelength radar beams capable of penetrating deep into the subsurface to assess the porosity profile, such as REASON onboard Europa Clipper \citep{Bayer2019} and RIME onboard JUICE \citep{Bruzzone2013}. 
Additionally, when considering the potential challenges for landers and rovers  \citep{Pappalardo2013, Hand2022} to operate on highly porous surfaces and when developing technologies like cryobots \citep{ValePereira2023} for exploring icy crusts and reaching subsurface oceans, it is imperative to account for this substantial compaction lengthscale.
Therefore, selecting target locations with low surface porosity would be of major interest in mission planning.

\section*{Statements and Declarations}
\subsection*{Funding and Competing Interests}

We acknowledge support from the ``Institut National des Sciences de l'Univers'' (INSU), the ``Centre National de la Recherche Scientifique'' (CNRS) and ``Centre National d'Etudes Spatiales'' (CNES) through the ``Programme National de Plan{\'e}tologie''. 

The authors have no competing interests to declare that are relevant to the content of this article.

\section*{Data Availability}
The authors assert that the data supporting the study's findings are included in the paper.

\appendix
\section{Appendix on the Rheology of Ice}
\label{sec:appendix_rheo}

Standard rheological models, for example the Burgers model \citep{Mellor1974}, have been used to study the mechanical behavior of snow. Nevertheless, as \citep{Mellor1974} pointed at, snow is a much more complicated material than bulk ice, and we currently do not have enough available data to define accurate general constitutive equations.
Hence we have to rely on simplifications, such as treating snow as a purely elastic \citep{Wilson1994} or purely viscous solid \citep{Nimmo2003}.

\subsection{Purely Elastic Solid}
\label{sec:appendix_ela}
For a a purely elastic solid, the strain is related to the applied stress using Hooke's Law:
\begin{equation}
    \epsilon = \frac{\sigma}{E}.
    \label{eq:eps_hooke}
\end{equation}
The strain rate can be expressed as the reduction of the pore radius $a$ as in \citep{Fowler1985}:
\begin{equation}
	\dot{\epsilon} = \frac{d a}{d t} \frac{1}{a} 
\end{equation}
Then, following \citep{Nimmo2003}, we can express the strain rate as a function of the porosity:
\begin{equation}
	\dot{\epsilon} = -\frac{d \phi}{d t} \frac{1}{\phi}. 
    \label{eq:strain_phi}
\end{equation}
Thus, taking the differential of Equation \eqref{eq:eps_hooke} and setting the stress equal to the overburden pressure $P(z)$ results in:
\begin{equation}
 d \epsilon = 	-d \phi \frac{1}{\phi} =  d P \frac{1}{E}
\end{equation}
where $E$ is the elastic Young modulus. Integrating from $z=0$ to $z$:
\begin{equation}
	\int_{\phi_0}^{\phi(z)} \frac{1}{\phi} d\phi = - \frac{1}{E} \int_0^{P(z)} dP(z)
\end{equation}
gives the expression of porosity that was assumed by Wilson \citep{Wilson1994}:
\begin{equation}
	\phi(z) = \phi_0 \exp\left(-\frac{1}{E}P(z) \right).
\end{equation}
This expression elucidates that the assumption of exponential decay of porosity with pressure  can be mathematically derived by assuming that the material is  purely elastic. 
Additionally, using this, we can establish  relationship between the compaction coefficient and the inverse of the Young modulus:
\begin{equation}
	\lambda = \frac{1}{E}.
\end{equation}

\subsection{Purely Viscous Solid}
\label{sec:appendix_visc}
For a purely viscous solid, the strain rate of the pores relationship to the stress is given by \citep{Mellor1974, Fowler1985}:
\begin{equation}
	\dot{\epsilon} = \frac{\sigma}{\eta}
\end{equation}
By taking the stress to be equal to the overburden pressure $P(z)$, and using \eqref{eq:strain_phi}, this leads to the differential equation found in \citep{Fowler1985, Nimmo2003}:
\begin{equation}
	\frac{d \phi}{d t}  = - \phi \frac{P(z)}{\eta}
\end{equation}
Leading to the expression of porosity given by Fowler \citep{Fowler1985}:
\begin{equation}
    \phi(z, t) = \phi_i(z) \exp\left(-\frac{t}{\eta} P(z)\right).
\end{equation}

%\bibliographystyle{sn-vancouver}
%\bibliography{library}

\end{document}